\documentclass[aps,prd,reprint,nofootinbib,showkeys,showpacs,longbibliography,floatfix,preprintnumbers]{revtex4-1}

\pdfoutput=1
\usepackage{amsmath}        
\usepackage{booktabs}       
\usepackage{array}          
\usepackage{multirow}       
\usepackage{graphicx}       
\usepackage{dcolumn}        
\usepackage{xspace}         
\usepackage{siunitx}        
\usepackage{amssymb}
\usepackage{bm}
\usepackage[american]{babel}
\usepackage{booktabs}
\usepackage[T1]{fontenc}  
\usepackage[]{hyperref}  
\usepackage[utf8]{inputenc}  
\usepackage{url}
\usepackage[dvipsnames]{xcolor}
\usepackage{tabularx}
\usepackage{booktabs} 

\DeclareUnicodeCharacter{2212}{-}

\makeatletter

\newcommand{\Rmnum}[1]{\expandafter\@slowromancap\romannumeral #1@}
\definecolor{greenW}{rgb}{0.0, 0.55, 0.1}

\newcommand{\Neff}{N_{\mathrm{eff}} }

\makeatother

\hypersetup{
	citecolor = red,
	linkcolor = blue,
	urlcolor = magenta
}

\begin{document}

\title{Extra Radiation Cosmologies: Implications of the Hubble Tension for eV-scale Neutrinos}

\author{Helena Garc\'ia Escudero}
\email{garciaeh@uci.edu}

\author{Kevork N.\ Abazajian}
\email{kevork@uci.edu}

\affiliation{Center for Cosmology, Department of Physics and Astronomy, University of California---Irvine, Irvine, CA 92697-4575, USA}

\preprint{UCI-HEP-TR-2025-21}

\begin{abstract}

We present a new analysis on sterile neutrino cosmologies using the Dark Energy Spectroscopic Instrument (DESI) second data release (DR2) baryon acoustic oscillation (BAO) measurements in combination with cosmic microwave background (CMB), CMB lensing, and supernova data. We show that BAO observables are intrinsically less sensitive to the combined effects of relativistic energy density, $\Neff$, and the sum of neutrino masses, $\Sigma m_\nu$, which are both augmented in sterile neutrino cosmologies. With SH0ES local expansion rate, $H_0$, data, we find $\Neff = 3.43 \pm 0.13$, reducing the Hubble tension to $2.4\sigma$. For a 0.1~eV sterile neutrino, we find $\Neff=3.50$ as the best fit.  For this representative $\Neff$, we find an upper limit of $m_s < 0.17$ eV (95\% CL), greater than a factor of four weaker than standard constraints on $\Sigma m_\nu$. When SH0ES is included, light sterile neutrinos with masses $m_s\simeq0.1$--$0.2$ eV are favored at $\gtrsim 3\sigma$, whereas eV-scale sterile masses remain strongly excluded by the data in the cosmologies we study.  Our findings confirm our previous results that partially thermalized sub-eV sterile neutrinos are preferred by the SH0ES $H_0$ data. The preferred $m_s$ mass scale overlaps with, but is not identical to, that favored in neutrino oscillation solutions to short-baseline anomalies. 
\end{abstract}


\maketitle

\section{Introduction}

Neutrinos play a fundamental role in cosmology, influencing several epochs, including but not limited to: the early Universe at weak freeze-out and primordial Big Bang nucleosynthesis (BBN), the growth of large-scale structure (LSS), as well as the anisotropies of the cosmic microwave background (CMB).  Their small but nonzero masses, inferred from oscillation experiments \citep{Super-Kamiokande:1998kpq,KamLAND:2002uet,SNO:2002tuh}, and their relativistic behavior in the early Universe leave measurable imprints on both the CMB and LSS. In the early Universe, neutrinos were all relativistic and contributed to the total radiation density, parameterized by the effective number of relativistic species, $N_{\rm eff}$. The standard thermal history predicts a value of $\Neff = 3.044$, slightly greater than the integer value due to heating of the plasma during $e^+/e^-$ annihilation \cite{Lesgourgues_Mangano_Miele_Pastor_2013, DeSalas:2018rby}. Deviations from this prediction would provide evidence of new physics such as additional light relics, non-standard neutrino interactions, or the presence of sterile neutrinos ($\nu_s$), in the case of $\Neff > 3.044$ \cite{Gerbino:2022nvz,Abazajian:2017tcc, Dolgov:2002wy}, or new early-Universe dynamics such as massive particle decay or low reheating temperature scenarios in the case of $\Neff <3.044$~\cite{Kawasaki:1999na,Kawasaki:2000en,Giudice:2000ex,Hasegawa:2019jsa,Hasegawa:2020ctq,Abazajian:2023reo,ACT:2025tim}.

It has been known for some time that finite neutrino masses suppress the growth of cosmic structures on scales smaller than the horizon at matter-radiation equality, due to a combination of free streaming of neutrinos below that scale and the scale-dependence of the linear growth of structure \cite{Lesgourgues:2006nd,Abazajian:2016hbv}. This characteristic suppression of clustering at smaller scales with increasing sum of neutrino masses was the pioneering method proposed for high sensitivity of cosmological LSS observables to the value of the total neutrino mass. The sum of neutrino masses here is $\Sigma m_\nu \equiv \Sigma_i m_i$, where $i$ sums over all thermalized mass eigenstates. Recently, the constraints on $\Sigma m_\nu$ have become so strong as to entertain the possibility of extra growth on smaller scales due to a fifth force in the dark matter sector, or other new physics, dubbed the ``negative neutrino mass'' problem \cite{Craig:2024tky}. More complicated models of reionization, altering the optical depth to the CMB, may also be responsible for this tension \cite{Jhaveri:2025neg}, as well as evolving dark energy models \cite{Elbers:2024sha}, which can alleviate constraints on $\Sigma m_\nu$ from LSS. Additional potential solutions are discussed in Ref.~\cite{Lynch:2025ine}.

The presence of extra relativistic energy density, whether in extra neutrino species or other relativistic species, produces a similar suppression of LSS below the horizon at matter-radiation equality due to the delay of this equality with increasing $\Neff$  \cite{Lesgourgues:2006nd,Abazajian:2016hbv}. This makes $\Sigma m_\nu$  and $\Neff$ positively correlated in their effects on LSS, with measurements of both being key goals of cosmological surveys \cite{Gerbino:2022nvz}.

Altering $\Neff$ modifies key observable features of the CMB. An increase in $\Neff$ reduces the angular size of the sound horizon through its effect on the expansion rate, scaling as $H(z)^{-1}$, and also alters Silk photon damping, which scales as $H(z)^{-1/2}$ as a scattering phenomenon. In early data from high multipole measures of the CMB found evidence for high $\Neff$ via this physics, with the Atacama Cosmology Telescope (ACT) finding approximately $2\sigma$ evidence for high $\Neff = 4.56 \pm 0.75$ \cite{Dunkley:2010ge} and the South Pole Telescope finding $\Neff = 3.86 \pm 0.42$ \cite{Keisler:2011aw}. This spurred interest in the possible evidence of light sterile neutrinos from cosmology, where it was found that the late time measure of the critical density in matter, $\Omega_m$, from Type Ia supernovae surveys was critical in inferring the preference or lack thereof of light sterile neutrinos in cosmology \cite{Joudaki:2012uk}. Subsequently, measurements of low amplitude of fluctuations at small scales, parameterized as $\sigma_8$ by their amplitude at $8 h\, \mathrm{Mpc}^{-1}$, renewed interest in cosmological preference for light sterile neutrinos  \cite{Battye:2013xqa,Wyman:2013lza,Dvorkin:2014lea}. Then, with the results from Planck 2018, massive neutrinos were very constrained by a combination of a low measurement of the optical depth to the CMB and high CMB lensing amplitude in that dataset, leaving extra sterile neutrinos highly constrained \cite{Planck:2018nkj}. 

The interplay between $N_{\rm eff}$ and $\Sigma m_\nu$ is particularly important in the context of current cosmological tensions, most notably the discrepancy between local distance-ladder measurements of the Hubble expansion rate, $H_0$, and the lower values inferred from the CMB under the $\Lambda$CDM model \cite{DiValentino:2021izs,CosmoVerseNetwork:2025alb,DiValentino:2025otz}. 
BAO data alone does not directly determine $H_0$, but when combined with other datasets such as BBN or CMB lensing, it generally prefers a slightly higher $H_0$ compared to the baseline Planck $\Lambda$CDM value, though still significantly lower than the local distance-ladder measurements~\cite{DESI:2025zgx}.
Increasing $N_{\rm eff}$ reconciles the CMB and BAO's inferred lower $H_0$ with the higher, locally measured, values of $H_0$ \cite{Bernal:2016gxb}. Increasing $\Neff$ reconciles the CMB inferred $H_0$ by decreasing the sound horizon at recombination by increasing the contribution to the expansion rate $H(z)$ through the radiation-dominated epoch into the era of matter-radiation equality:
\begin{equation}
r_\mathrm{s} (z_\mathrm{rec}) = \int^{z_\mathrm{rec}}_\infty {\frac{c_s\, dz}{H(z)}}\, ,
\label{eq:sound_horizon}
\end{equation}
where $r_\mathrm{s}$ is the sound horizon at recombination, $z_\mathrm{rec}$ is the redshift of recombination, and $c_s$ is the sound speed of the plasma. It has been established that $\Neff$ is as robust of a solution to the Hubble tension as other new physics, including early dark energy, given statistical tests of the data sets \cite{Escudero:2022rbq}, with a preference for a high value of $\Neff \approx 3.5$ \cite{Escudero:2024uea}.

Extra neutrino density from cosmology begs the question as to its relation with hints from short-baseline neutrino oscillation experiments that suggest the possible existence  of additional neutrino species beyond the three active flavors \cite{LSND:2001aii,MiniBooNE:2010idf}. Fits to short baseline oscillation anomalies prefer $\nu_s$, with masses ($m_s$) at the eV scale \cite{Dasgupta:2021ies,Norman:2024hki}. Unlike active neutrinos, sterile neutrinos do not participate in weak interactions, but, if partially or completely thermalized in the early Universe, they can contribute to both $\Sigma m_\nu$ and  $N_\mathrm{eff}$ \cite{Abazajian:2017tcc}. The partial or full thermalized presence of $\nu_s$ alters cosmological observables similar to massive active neutrinos and extra relativistic energy density, which are typically tested individually as deviations from minimal $\Lambda$CDM. The presence of an extra $\nu_s$ is not identical to the individual parameters, as the extra species carries relativistic energy density and neutrino mass as a distinct, massive neutrino state that affects cosmological observables uniquely.

In exploring these scenarios, we found in Ref.~\cite{Escudero:2024uea} that cosmologies with a partially-thermalized sterile neutrino component are strongly favored when incorporating the SH0ES measurement of the Hubble constant, when compared to standard $\Lambda$CDM. This preference arises because a sterile neutrino increases $N_{\rm eff}$, thereby reducing the sound horizon and reconciling the CMB-inferred expansion rate with SH0ES. Because of the relative insensitivity of BAO observables to increasing $\Sigma m_\nu$ when increasing $\Neff$, we found that even massive, partially thermalized sterile neutrinos with $m_s = 0.1 \,\mathrm{eV}$ are preferred at $3.3\sigma$ relative to standard $\Lambda$CDM. The fact that such models are preferred by the data provides the strong motivation for our present work, where we systematically investigate the nature of BAO, LSS, and CMB sensitivity to massive sterile neutrinos in light of current observational constraints.  Our work aims to test the status of current cosmological data on the presence of partially to fully thermalized sterile neutrinos, as the presence of massive neutrinos is constrained by LSS, while the Hubble tension prefers extra relativistic energy density.

The structure of this manuscript is as follows. In the next section, \S\ref{sec:implicat}, we review the effects on cosmological observables by massive neutrinos and extra relativistic energy density, both of which can be provided by $\nu_s$. We connect this with the implications of the Hubble tension's shift on $\Neff$ and inferred limits on $m_s$. In \S\ref{sec:datasets}, we present tests of sterile neutrino scenarios with recent cosmological data sets. Our results are presented and analyzed in \S\ref{sec:results}. We conclude with a discussion of future prospects in Section \ref{sec:discussion}.

\section{Neutrino Cosmology: $\pmb{\Neff}$ \& $\pmb{\Sigma m_\nu}$ from CMB, LSS, and BAO}
\label{sec:implicat}

Massive neutrinos and their related relativistic energy density play a fundamental role in cosmology, leaving distinctive imprints on the CMB, LSS, and BAO observables. We review these signatures and emphasize how they complement each other in the presence of additional massive neutrinos. The effects of extra, massive neutrinos are well known in the case of their impact on LSS, which we review for completeness. Their effects on BAO observables is less well known, which we expand on more as a result. 

\subsection{Suppression of LSS by  $\pmb{\Sigma m_\nu}$ \& $\pmb{\Neff}$ }

 A well-established effect in cosmology is the suppression of LSS power at relatively smaller scales induced by the free-streaming of massive neutrinos. The amplitude of this suppression is related to the total neutrino mass, $\Sigma m_\nu$, and arises because massive neutrinos, becoming non-relativistic at late times, do not cluster efficiently when they are relativistic, and also suppress growth below their free streaming scale \citep{Eisenstein:1997ik,Hu:1997mj}. The fractional suppression in the matter power spectrum can be approximated as
\begin{equation}
\frac{\Delta P}{P} \simeq -8 f_\nu \simeq  -8 \frac{\Omega_\nu}{\Omega_m} \simeq  -\frac{1}{\Omega_m h^2} \frac{\Sigma m_{\nu}}{11.6\,\mathrm{eV}} \, , \label{eq:hueqn}
\end{equation}
where $f_\nu =\Omega_\nu/\Omega_m$ is the neutrino fraction of the matter density~\citep{Hu:1997mj}. The neutrino density is related to the total mass by $\Omega_{\nu} = {\Sigma m_{\nu}}/{93.13 \, h^2 \,\mathrm{eV}}$, with $h$ the Hubble parameter in units of $100\,\mathrm{km\,s^{-1}\,Mpc^{-1}}$. As can be readily seen from Eq.~\eqref{eq:hueqn}, a sensitivity of 1\% in change in power, $\Delta P/P$, will result in a sensitivity to $\Sigma m_\nu$ of 10 meV, given $\Omega_m h^2 \approx 0.1$. This suppression is most significant on scales that enter the horizon prior to neutrinos becoming non-relativistic, which coincides with their free-streaming scale, suppressing growth of structure, leading to a reduced amplitude of matter clustering at below the horizon size at matter-radiation equality. The often used metric of fluctuation amplitudes at the $8 h \mathrm{Mpc}^{-1}$ scale is $\sigma_8$, which is below that horizon size is the amplitude of fluctuation. Therefore, $\sigma_8$ is a useful measure of the effects of neutrino mass suppression at these cosmologically smaller scales. 

The effective number of relativistic species, $N_{\rm eff}$, alters structure growth through a different mechanism. Increasing $N_{\rm eff}$ raises the radiation energy density, increasing the expansion rate during radiation domination, and delaying matter--radiation equality. When fixing the amplitude of scalar perturbations from inflation, this delay enhances the decay of perturbations in the radiation domination era, and shifts the horizon size of matter radiation equality to larger scales. This results, as in the case of increasing the mass of neutrinos, as a decrease in $\sigma_8$.

\begin{figure*}[t]
  \centering
  \begin{minipage}[t]{0.49\textwidth}
    \centering
    \includegraphics[width=\linewidth]{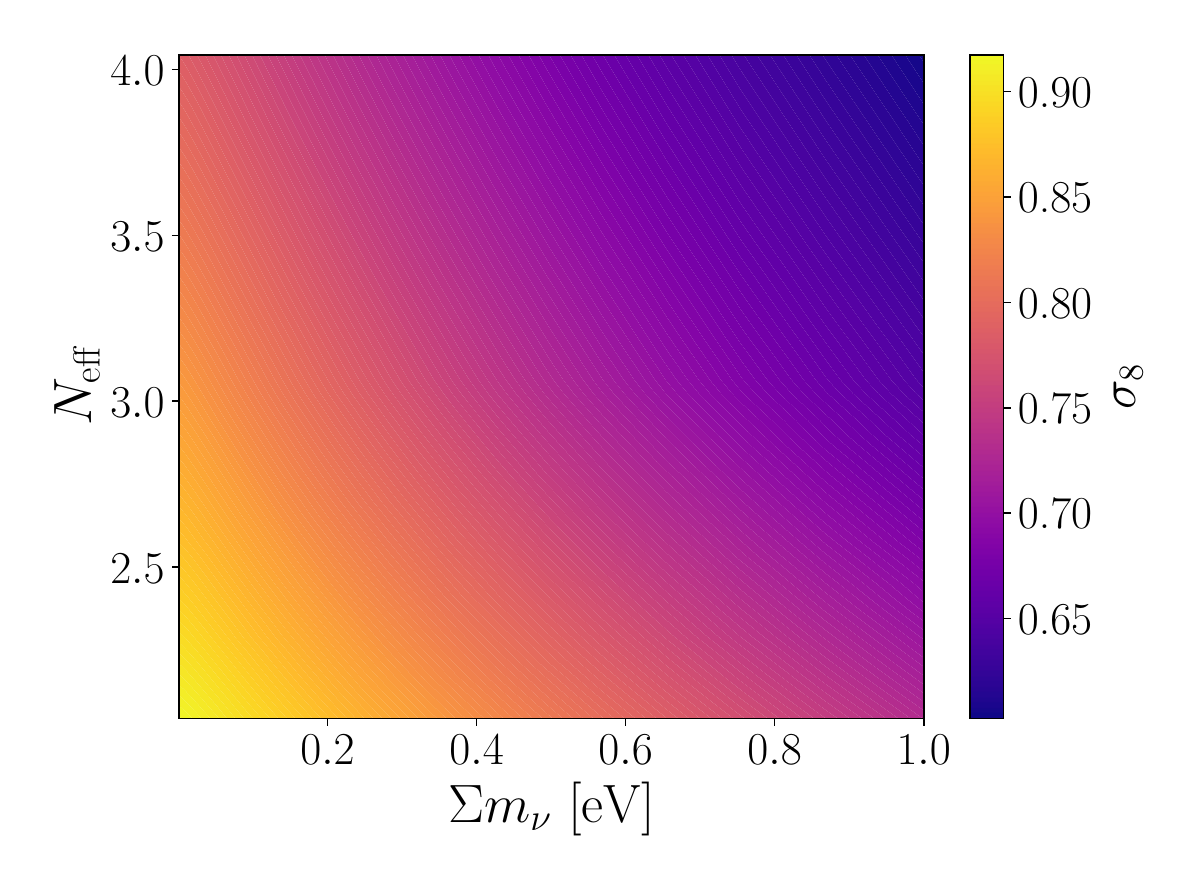}
  \end{minipage}\hfill
  \begin{minipage}[t]{0.49\textwidth}
    \centering
    \includegraphics[width=\linewidth]{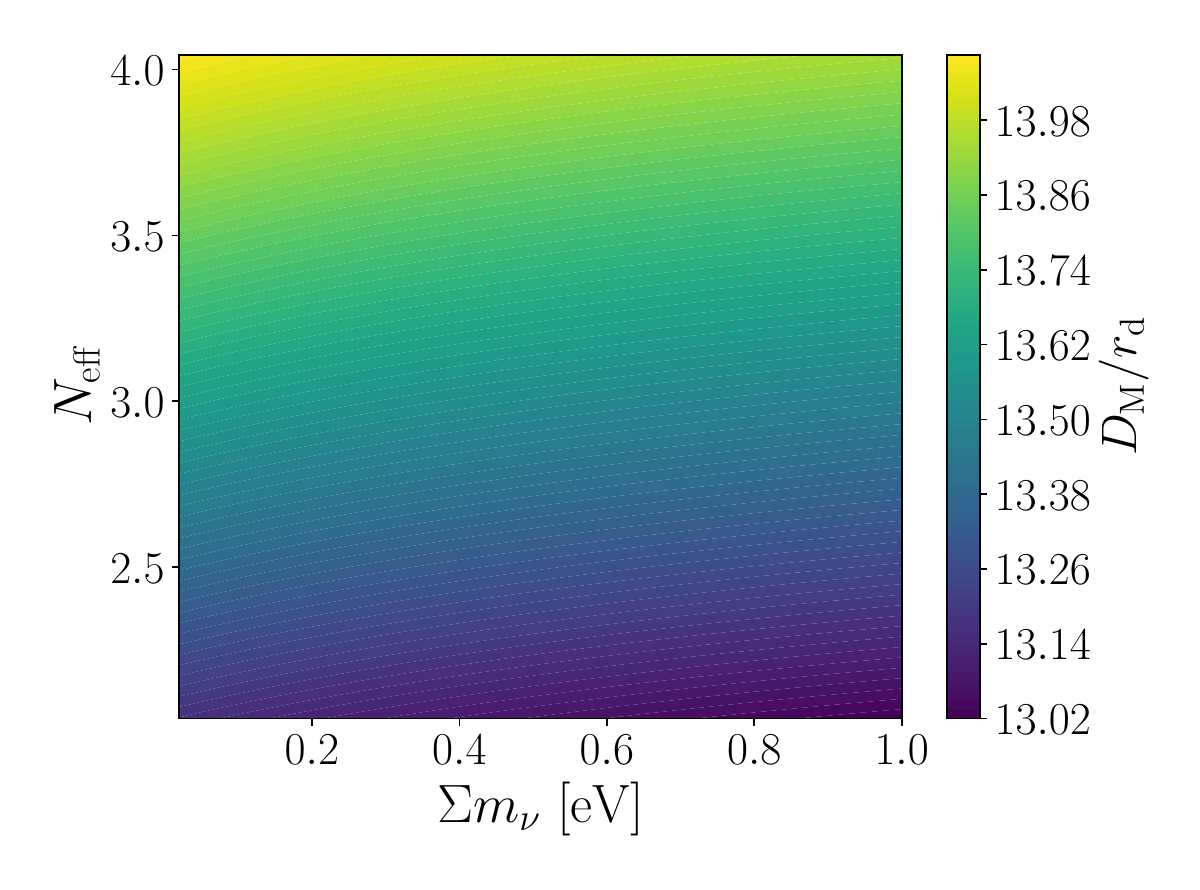}
  \end{minipage}
  \caption{\textbf{Left panel:} contours of the clustering amplitude $\sigma_8$ as a function of the total neutrino mass $\Sigma m_\nu$ and the effective number of relativistic species $N_{\mathrm{eff}}$, computed while keeping the remaining $\Lambda$CDM parameters fixed at their standard values. Lowering $\Sigma m_\nu$ and $N_{\mathrm{eff}}$ increases $\sigma_8$, while raising them decreases it. \textbf{Right panel:} contours of the BAO distance ratio $D_\mathrm{M}/r_\mathrm{d}$ at $z=0.51$ for different $\Sigma m_\nu$ and $N_{\mathrm{eff}}$ values when the total matter density is fixed. An increase in neutrino mass is compensated by a reduction in the cold dark matter fraction. These contours reveal the anticorrelated response of $\Sigma m_\nu$ and $N_{\mathrm{eff}}$ through their opposite effects on the $D_\mathrm{M}/r_\mathrm{d}$ BAO observable. Together, these panels highlight the complementary ways in which neutrino properties imprint on growth and geometry.}
  \label{fig:s8}
\end{figure*}

Increases in both $\Sigma m_\nu$ and $N_{\rm eff}$ lower $\sigma_8$, suppressing the amount of structure. Therefore, they are correlated parameters when analyzing their effects in LSS observables, including galaxy power spectra and CMB lensing. The left panel in Figure~\ref{fig:s8} shows this correlation between $\Sigma m_\nu$ and $ N_{\rm eff}$  where the contours of constant $\sigma_8$ trace diagonal directions in the $\Sigma m_\nu - N_{\rm eff}$ plane. While the physical origins differ, neutrino free-streaming versus delayed matter-radiation equality, the net imprint on late-time clustering is similar.

\subsection{BAO: Geometry and Anticorrelation}

BAO arise from pressure waves in the photon--baryon plasma of the early Universe. Measurements of the BAO feature were first evidenced in the Sloan Digital Sky Survey (SDSS) \cite{Eisenstein:2005su} \& the Two-Degree Field Galaxy Redshift Survey \cite{2dFGRS:2005yhx}, and more accurately measured in the Baryon Oscillation Spectroscopic Survey~\citep{BOSS:2016wmc}, and the Extended Baryon Oscillation Spectroscopic Survey~\citep{eBOSS:2020yzd}, with the Dark Energy Spectroscopic Instrument (DESI) providing the most recent  measurements~\citep{DESI:2025zgx}. BAO observations enable precise determinations of cosmological distances and the expansion history. In a spatially flat cosmological model with matter, radiation, and a cosmological constant, the Hubble parameter at redshift $z$ is given by the Friedmann Equation as
$H(z) = H_0 \sqrt{ \Omega_m (1 + z)^3 + \Omega_r (1 + z)^4 + 1 - \Omega_m}$,
where $H_0$ is the present-day Hubble constant, $\Omega_m$ the matter density relative to the critical density today, and $\Omega_r$ the commensurate radiation density. BAO observations do not measure distances directly, but angles and comoving distance ratios. The relevant distance measures are the comoving angular diameter distance $D_\mathrm{M}(z)$ and comoving line-of-sight distance $D_\mathrm{H}(z)$, given by
\begin{align}
D_\mathrm{M}(z) &= \int_0^z \frac{c \, dz'}{H(z')} ,\\
D_\mathrm{H}(z) &= \frac{c}{H(z)}.
\end{align}
BAO observations constrain the ratio of these distances to the sound horizon at the drag epoch $r_\mathrm{d}$, which corresponds to the comoving distance traveled by acoustic waves in the photon–baryon fluid before baryon decoupling. Evaluating the sound-horizon integral in Eq.~\eqref{eq:sound_horizon} at the drag epoch, $z_{\rm d}$, rather than at recombination, provides $r_\mathrm{d}$. The transverse BAO scale measures the angular size of $r_\mathrm{d}$ and is given by $D_\mathrm{M}(z)/r_\mathrm{d}$, while the radial BAO scale probes the distance along the line of sight through $D_\mathrm{H}(z)/r_\mathrm{d}$.  Together, these measurements provide powerful probes of the expansion history.

The geometric observables of angular diameter distance and comoving distance provide an additional channel for sensitivity to neutrino properties. The effect of increasing $\Sigma m_\nu$ and $N_\mathrm{eff}$ not only suppresses the growth of structure at smaller scales, but also alters the background expansion rate, effectively changing the cosmic distances inferred from BAO.

To delve into the effects of neutrino properties, we consider two well-separated redshifts in cosmic history: $z=0.51$, corresponding to the mean first bin  redshift of luminous red galaxies (LRG) in DESI Second Data Release (DR2), and $z=2.33$, corresponding to the DESI DR2 mean redshift for Ly$\alpha$ forest. Considering $z=0.51$ first, there is an anticorrelation that can be understood in terms of how neutrino properties shift the BAO peak in the power spectrum. Neutrinos with combined masses of at least $58$~meV affect the BAO feature in non-trivial ways. As described by Thepsuriya and Lewis~\cite{Thepsuriya:2014zda}, increasing $\Sigma m_\nu$ alone tends to shift the BAO peak down and to smaller spatial scales because the sound horizon shrinks as more neutrino mass augments $\Omega_m$, during and after matter–radiation equality, increasing $H(z)$ in the drag epoch, and therefore reducing the integrand in Eq.~\eqref{eq:sound_horizon}. However, if the total matter density is held fixed, then a larger neutrino mass requires a reduction in the cold dark matter fraction, which instead shifts the BAO peak to larger spatial scales, because neutrinos are just starting to act as matter (becoming nonrelativistic) around matter-radiation equality. In contrast, increasing $N_\mathrm{eff}$ raises the radiation energy density, delays matter–radiation equality, and shifts the BAO scale to smaller spatial scales by also reducing the integrand in Eq.~\eqref{eq:sound_horizon}.  As we describe below, $D_\mathrm{M}$ at a low redshift of $z=0.51$ is not significantly affected when $\Sigma m_\nu$ is varied, and it is $r_\mathrm{d}$ that largely determines the behavior of the BAO measurable, $D_\mathrm{M}/r_\mathrm{d}$. Therefore, the opposite response of the ratio to $\Sigma m_\nu$ and $\Neff$ are directly reflected in the anticorrelation pattern shown in the right panel of Fig.~\ref{fig:s8}.

\begin{figure*}[t]
    \centering
    \includegraphics[width=0.95\linewidth]{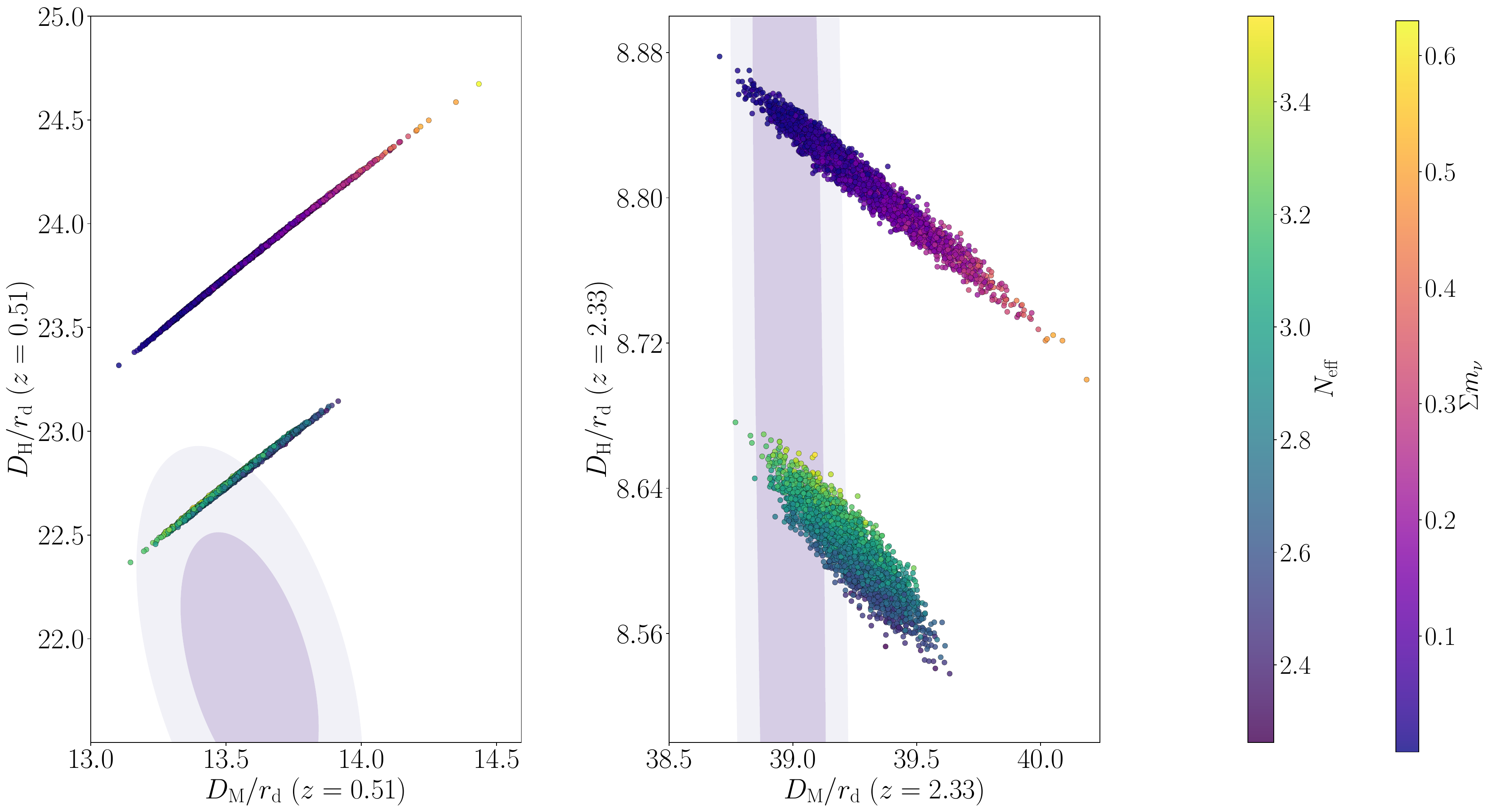}\\
    \ \\
    \includegraphics[width=0.95\linewidth]{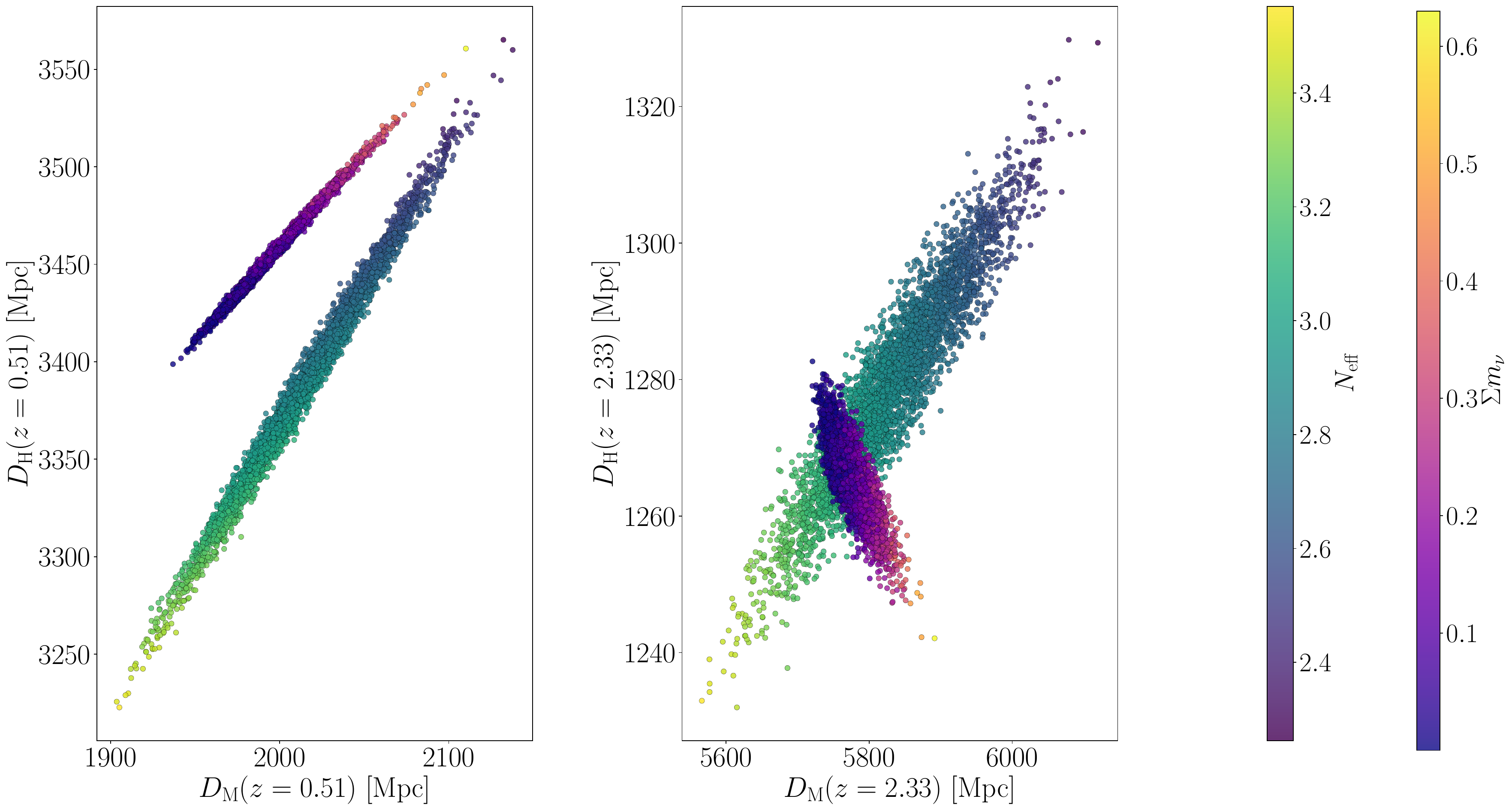}

        \caption{\textbf{Top Panels:} $D_\mathrm{M}/r_\mathrm{d}$ shown along the horizontal axis and $D_\mathrm{H}/r_\mathrm{d}$ along the vertical axis, calculated from the parameters of the MCMC chains from the \textit{Planck} 2018 CMB likelihoods alone, for two distinct models with free neutrino mass and $N_\mathrm{eff}$, separately. The left panel corresponds to a redshift of 0.51 and the right panel to a redshift of 2.33. The colors indicate the values of $N_\mathrm{eff}$ and  $\Sigma m_\nu$ specified in the color bar. A positive vertical offset has been applied to the $\Sigma m_\nu$ chains in both panels to ensure clarity of the figure ($z=0.51$ vertical offset of $+1$ ; $z=2.33$ vertical offset of $+0.2$). The purple contours denote the $1\sigma$ and $2\sigma$ DESI DR2 BAO measurements of $D_\mathrm{M}/r_\mathrm{d}$ and $D_\mathrm{H}/r_\mathrm{d}$. \textbf{Bottom Panels:} To show their dependence on the neutrino parameters independent of $r_\mathrm{d}$, $D_\mathrm{M}$ is shown along the horizontal axis and $D_\mathrm{H}$ is along the vertical axis for models with free neutrino mass and $N_\mathrm{eff}$, including only the \textit{Planck} 2018 CMB likelihoods. A positive vertical offset of +100 Mpc has been applied to the $\Sigma m_\nu$ chain points at $z=0.51$ to ensure clarity of the figure.}
    \label{fig:aaf_DM_DH_rd_panels}
\end{figure*}

BAO measurements are predicated on the high amount of cosmological information from the CMB. Therefore, we analyze the distribution of Markov Chain Monte Carlo (MCMC) chain points preferred by the CMB data in the space of the BAO observables, as shown in Fig.~\ref{fig:aaf_DM_DH_rd_panels}, which plots a subset of the MCMC chain points from the \textit{Planck} 2018 only likelihoods. We choose cosmologies with one extra parameter---either $\Neff$ or $\Sigma m_\nu$, to show their individual effects on BAO observables. There, one can see that increasing $N_\mathrm{eff}$ leads to a {\em decrease} in $D_\mathrm{H}/r_\mathrm{d}$ at redshift $z=0.51$ and an {\em increase} in that ratio at redshift $z=2.33$. The opposite happens when $\Sigma m_\nu$ is increased: $D_\mathrm{H}/r_\mathrm{d}$ {\em increases} at redshift $z=0.51$ and {\em decreases} at redshift $z=2.33$.  For $D_\mathrm{M}/r_\mathrm{d}$, larger $N_\mathrm{eff}$ results in an {\em decrease} of this ratio at both redshifts $z=0.51$ and $z=2.33$ and larger $\Sigma m_\nu$ results in an {\em increase} of this ratio at both redshifts $z=0.51$ and $z=2.33$, as shown in Fig.~\ref{fig:aaf_DM_DH_rd_panels}. The end result of these trends is an {\em anticorrelation} between $N_\mathrm{eff}$ and $\Sigma m_\nu$ when we look at BAO observables. This was previously described in the more constrained parameter space in Fig.~\ref{fig:s8}. In short, a simultaneous increase in both $N_\mathrm{eff}$ and $\Sigma m_\nu$ shifts $D_\mathrm{M}/r_\mathrm{d}$ and $D_\mathrm{H}/r_\mathrm{d}$ in opposite directions, allowing for the predicted BAO measurement to stay constant, and consistent with the observed values. That is, in the case if partially or fully thermalized massive sterile neutrinos, BAO measurements can accommodate the simultaneous presence of increased $N_\mathrm{eff}$ and $\Sigma m_\nu$.

\begin{figure*}[t]
    \centering
    \begin{minipage}{0.49\textwidth}
        \centering
        \includegraphics[width=\linewidth]{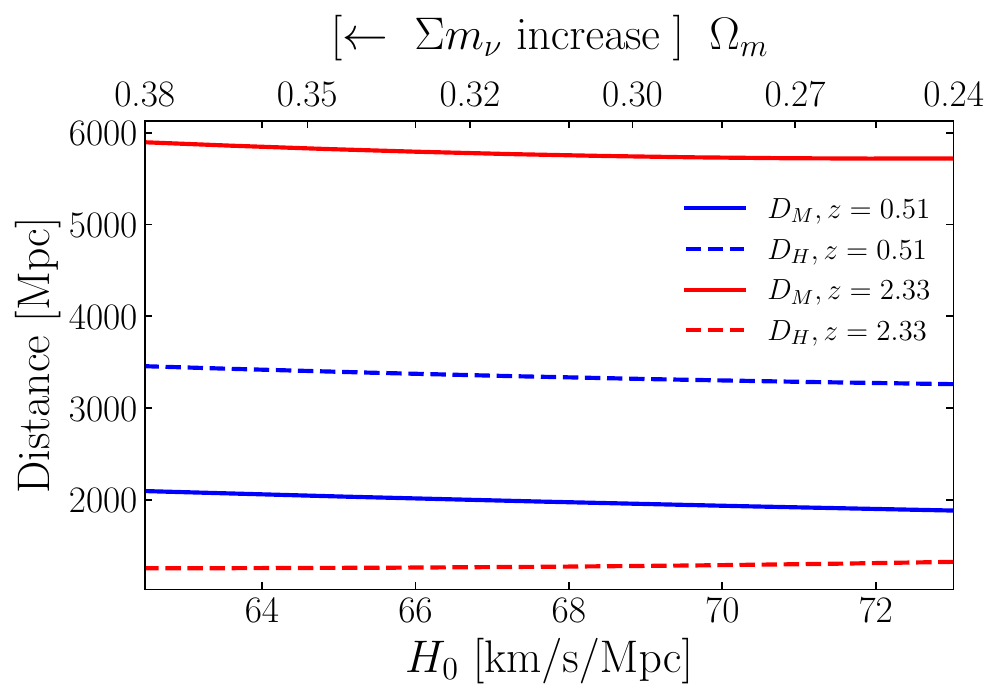}
    \end{minipage}
    \hfill
    \begin{minipage}{0.49\textwidth}
        \centering
        \includegraphics[width=\linewidth]{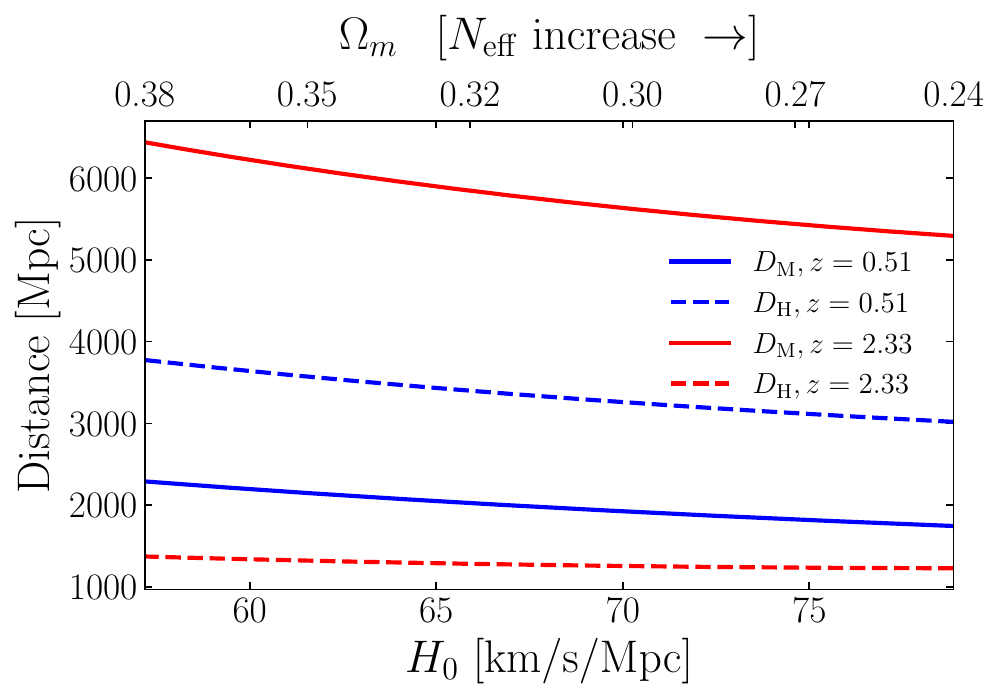}
        \label{fig:aaf_nnu_linear_combined.pdf}
    \end{minipage}
   \caption{
      Variation of the transverse distance $D_\mathrm{M}$ (solid lines) and the radial distance $D_\mathrm{H}$ (dashed lines) is shown at two redshifts: $z=0.51$ (blue) and $z=2.33$ (red). These are plotted as the Hubble constant $H_0$ along the lower horizontal axis and the matter density $\Omega_m$ are varied along their linear degeneracy from the \textit{Planck} 2018 CMB likelihoods, as described in the text. The left panel shows models where $\Sigma m_\nu$ increases in correlation with $\Omega_m$ (to the left) and the right panel shows models where $N_\mathrm{eff}$ increases in anticorrelation to $\Omega_m$, to the right.}
    \label{fig:nnu_mnu_linear_panels}
\end{figure*}

The change in slope of the CMB-derived points between the left and right panels of Fig.~\ref{fig:aaf_DM_DH_rd_panels} is driven by the opposite responses of $D_\mathrm{H}/r_\mathrm{d}$  to variations in $\Sigma m_\nu$ and $N_\mathrm{eff}$. For increasing neutrino mass, $D_\mathrm{H}/r_\mathrm{d}$ increases at $z=0.51$ but decreases at $z=2.33$, while for increasing $N_\mathrm{eff}$ the trend is reversed: $D_\mathrm{H}/r_\mathrm{d}$ decreases at $z=0.51$ and increases at $z=2.33$. In the case of neutrino mass, this behavior arises from the degeneracy with $\Omega_m$, enforced by the CMB data: as $\Sigma m_\nu$ increases, the degeneracy with $\Omega_m$ requires it to rapidly increase to preserve the angular size of the sound horizon on the sky, as discussed in Ref.~\cite{Loverde:2024nfi}.  We fit the degeneracy between $\Omega_m$ and $H_0$ in the full likelihood to find the linear relation between the degeneracy of these parameters. Given this combination of effects, an increase in $\Omega_m$ drives a reduction of $D_\mathrm{H}$ at high redshift (the dashed red line in left panel of Fig.~\ref{fig:nnu_mnu_linear_panels}), while driving an increase in $D_\mathrm{M}$ at high redshift. For the low redshift case, $D_\mathrm{H}$ and $D_\mathrm{M}$ instead both increase with $\Sigma m_\nu$. This is why the contours invert in direction from the left to the right panel of Fig.~\ref{fig:aaf_DM_DH_rd_panels}, for the case of $\Sigma m_\nu$. 

For varying $N_\mathrm{eff}$, shown in the right panel of Fig.~\ref{fig:nnu_mnu_linear_panels}, $D_\mathrm{H}$ decreases as $N_\mathrm{eff}$ increases for both redshifts, which is shown in the bottom panels of Fig.~\ref{fig:aaf_DM_DH_rd_panels}. As previously discussed, $r_\mathrm{d}$ also decreases with increasing $\Neff$, but the fractional change in $r_\mathrm{d}$ is larger than that in $D_\mathrm{H}$. This dominance of the variation in $r_\mathrm{d}$ over that in $D_\mathrm{H}$ is the key factor responsible for the reversal of the slope of the CMB points across the two panels for {\em that} observable parameter, shown between the panels of Fig.~\ref{fig:aaf_DM_DH_rd_panels}. Though the contours invert for $\Sigma m_\nu$ and $\Neff$ at higher $z$ for different reasons, they preserve the anticorrelation between them seen at low $z$. Therefore, BAO observables accommodate a simultaneous increase in $\Sigma m_\nu$ and $\Neff$ across a wide range of redshifts.

Many of parts of these effects are explored in previous studies. In \citet{Loverde:2024nfi}, they show that BAO observables are sensitive to matter \emph{fractions} rather than absolute densities, and that within $\Lambda$CDM, the total matter fraction $\Omega_m$ must increase significantly as $\Sigma m_\nu$ grows, making geometric measurements a  powerful and complementary probe of neutrino mass alongside structure growth suppression. The effects of neutrino physics on the drag scale were also studied in \citet{Thepsuriya:2014zda}.
The relation between the BAO scale at different redshifts and its dependence on neutrino mass and number as constrained by the CMB was also explored in \citet{BOSS:2014hhw}, where the anticorrelation we discuss was also shown. In this work, we have gone beyond and shown explicitly where these relations in BAO observables arise in their sensitivities to $\Sigma m_\nu$ and $\Neff$.

\section{Cosmological datasets \& Methodology}
\label{sec:datasets}
\subsection{Baseline cosmological dataset}
For our baseline data, we use a combination of cosmological datasets to constrain our models. The datasets are similar to our previous work \cite{Escudero:2024uea}, but here we include the more newly available DESI DR2 data. \\

\noindent \textbf{CMB Planck Final Data Release (P18)}\\
The Planck 2018 (PR3) legacy data release~\cite{Planck:2018vyg,Planck:2019nip} provides highly precise measurements of the CMB temperature and polarization power spectra, with five out of six $\Lambda$CDM parameters determined to better than 1\% and the angular sound horizon measured to 0.03\%. We include high-$\ell$ \texttt{Plik} likelihoods, low-$\ell$ \texttt{commander} likelihoods for TT in $2 \leq \ell \leq 29$, \texttt{SimAll} for EE in the same range, and TE and EE spectra at $30 \leq \ell \leq 1996$.  \\

\noindent \textbf{CMB Lensing} \\ 
We incorporate CMB lensing measurements from the Planck NPIPE PR4 reconstruction~\cite{Carron:2022eyg}, which includes roughly 8\% more data than PR3 with improved processing, as well as ACT DR6 lensing data covering over 9,400 deg$^2$ of the sky~\cite{ACT:2023dou,ACT:2023kun}, achieving 2.3\% precision on the lensing power spectrum.\\  

\noindent \textbf{BAO: DESI DR2}  \\
We use BAO measurements from over 14 million galaxies and quasars in  DESI DR2. We adopt the DESI DR2 BAO likelihood~\citep{DESI:2025zgx}, which updates the previous DR1 measurements and includes the volume-averaged distance $D_V/r_\mathrm{d}$ from the Bright Galaxy Survey sample ($0.1<z<0.4$), $D_\mathrm{M}/r_\mathrm{d}$ and $D_\mathrm{H}/r_\mathrm{d}$ for LRG samples ($0.4<z<0.6$ and $0.6<z<0.8$), a combined LRG plus Emission Line Galaxies (ELG) tracers ($0.8<z<1.1$), ELG measurements ($1.1<z<1.6$), quasar clustering ($0.8<z<2.1$), and Ly$\alpha$ BAO constraints ($1.8<z<4.2$). We refer to this combination as ``DESI2.''  \\

\noindent \textbf{Type Ia Supernovae}  \\
Type Ia supernovae (SNe) provide an independent probe of the expansion history. We adopt the Pantheon+ sample~\cite{Scolnic:2021amr,Brout:2022vxf}, containing 1550 spectroscopically confirmed SNe Ia over $0.001 < z < 2.26$ with improved calibration and systematics relative to the previous Pantheon release~\cite{Pan-STARRS1:2017jku}.  \\

\noindent \textbf{Local Expansion Rate: $H_0$ (SH0ES)}  \\
We adopt the Supernovae and $H_0$ for the Equation of State of dark energy (SH0ES) measurement of $H_0$ for the local Hubble expansion rate. We use the Gaussian likelihood for the Hubble constant from the SH0ES collaboration~\cite{Riess:2021jrx}, $H_0 = 73.04 \pm 1.04\, \mathrm{km}\,\mathrm{s}^{-1}\,\mathrm{Mpc}^{-1}$. We introduce a $H_0$ prior directly in the likelihood rather than the absolute magnitude $M_b$ since using the latter only substantially affects models with non-standard late-time expansion histories, which are not considered in this work.  \\

\subsection{Complementary Data: SDSS DR12 BAO \& Full-Shape}
Since the LSS information from the full-shape power spectrum of DESI DR2 is not publicly available, we consider constraints from the latest BAO and full-shape analysis of SDSS DR12, which we dub SDSSBAOFS~\cite{BOSS:2016wmc}. Unlike BAO-only analyses, the full-shape likelihood is sensitive not only to the position of the acoustic feature but also to the broadband shape of the LSS clustering spectrum, which carries information about the underlying matter content and growth of structure as a function of scale. This makes it valuable for constraining the effects of massive neutrinos and extra $\Neff$, which both induce the characteristic suppression of the full-shape power at smaller scales.

\subsection{Methodology}

As we performed our analyses of these models given the cosmological data, we found a dependence on the final result for signals of massive, extra neutrinos that depend on whether one adopts Bayesian or frequentist statistics. Bayesian inference offers a systematic framework for model comparison through the evaluation of the Bayesian evidence \cite{Trotta:2008qt}, whereas frequentist approaches such as profile likelihoods provide complementary, prior-independent assessments of the parameter space \cite{Cowan:1998ji}. Frequentist techniques for parameter inference that employ profile likelihoods have historically been used less often in cosmology, largely because of their computational expense. However, they have recently gained traction due to their ability to yield results that complement Bayesian analyses. In the Bayesian framework, credible intervals depend on the specification of priors for all parameters and incorporate nuisance parameters through integration (i.e., marginalization). In contrast, frequentist confidence intervals do not inherently require priors and instead address nuisance parameters by maximizing the likelihood. These methodological differences, involving both prior dependence and the treatment of nuisance parameters, are especially relevant in scenarios where the data weakly constrain the model and/or when parameters are near physical boundaries. In Bayesian studies, this situation can give rise to effects driven by the choice or volume of the prior (see, e.g., \cite{Gariazzo:2018pei,Smith:2020rxx,Herold:2021ksg,Herold:2022iib,Holm:2023laa} for applications in cosmology). Importantly, both Bayesian and frequentist approaches converge in the limit of sufficiently large datasets. 

Since we are interested in extra, massive sterile neutrino models, we implement the extra relativistic energy density above the standard energy content in the active neutrinos, which we adopt to be $N_\nu \equiv 3.044$. We model such a cosmology as a partially thermalized sterile neutrino model with $\Sigma m_\nu$ and $\Delta\Neff\equiv \Neff - N_\nu$ as additional  parameters  to the six $\Lambda$CDM ones \cite{Escudero:2024uea}, which is the formulation required by \texttt{CAMB} \cite{Lewis:1999bs}:
\begin{equation}
m^{\textrm{eff}}_{\textrm{s}} = \Delta N_{\textrm{eff}}~ m_{\textrm{s}}^\mathrm{ph}; \qquad \Delta N_{\textrm{eff}} =\beta.
\end{equation}
Importantly, $m^{\textrm{eff}}_{\textrm{s}}$ is not to be mistaken to be the physical mass. \texttt{CAMB} requires $m^{\textrm{eff}}_{\textrm{s}}$ and $\beta$ to be the free parameters describing the sterile neutrino. Here, $m_{\textrm{s}}^\mathrm{ph}$ denotes the physical mass of the additional sterile neutrino species only, excluding the three active Standard Model neutrinos. In this case, one encounters the model space where $\beta \sim 0.01$ and the physical mass approaches $m_{\textrm{s}}^\mathrm{ph}\sim 1\,\mathrm{keV}$, which is when the sterile neutrino can displace the need for CDM and becomes the (warm) dark matter. Because the sterile neutrino mass can be arbitrarily large and $\Delta\Neff$ arbitrarily small for a fixed contribution to the dark matter density, one must map the parameter space to a finite volume in the case of Bayesian statistics. In our results, we found that the often-used parameterization in $1/m_s$ leads to inaccurate results on constraints on massive, extra neutrinos, with the results depending on the statistical approach, as discussed further in the Appendix. Therefore, for our main results, we adopt either frequentist statistics or Bayesian models where only $m^{\textrm{eff}}_{\textrm{s}}$ or $\beta$ were individually left free. 

We employ frequentist approaches based on profile likelihoods, which provide a prior-independent way to assess constraints on cosmological parameters, including those bounded by physical limits. In this method, the likelihood is evaluated at fixed values of the parameter of interest while maximizing over all remaining cosmological and nuisance parameters. We employ 180 extremizations of the likelihood using the extremizing sampler provided in \texttt{Cobaya}~\cite{Torrado:2020dgo}. in order to best determine the likelihood's true value, due to the considerations we discussed in Ref.~\cite{Escudero:2024uea}. The general construction of confidence intervals for bounded parameters like $m_s$ was formalized in the Feldman--Cousins framework~\cite{Feldman:1997qc}, which defines coverage in the ``true parameter--measured parameter'' space. An equivalent description in terms of the likelihood-ratio test statistic has been outlined in Ref.~\cite{Herold:2024enb}.

We also perform Bayesian inference using the publicly available \texttt{Cobaya} package, implementing the MCMC sampler~\cite{Lewis:2002ah,Lewis:2013hha} with fast dragging~\cite{2005math......2099N}, and computing theoretical predictions with the \texttt{CAMB} cosmological Boltzmann solver~\cite{Lewis:1999bs,Howlett:2012mh}. Each analysis is run with four parallel MCMC chains, and convergence is assessed using the Gelman-Rubin statistic, requiring $R-1 < 0.01$ for all parameters.

We adopt flat priors for all cosmological parameters, chosen to extend well beyond the region of significant likelihood support. Specifically, we vary the baryon density $\Omega_\mathrm{b} h^2$ from $0.005$ to $0.1$, the cold dark matter density $\Omega_\mathrm{c} h^2$ from $0.01$ to $0.99$, the scalar amplitude $\log(10^{10} A_\mathrm{s})$ from $1.61$ to $3.91$, the scalar spectral index $n_\mathrm{s}$ from $0.8$ to $1.2$, the reionization optical depth $\tau_{\rm reio}$ from $0.01$ to $0.8$, and the angular scale of the sound horizon $100\theta_\mathrm{MC}$ from $0.5$ to $10$. For the neutrino sector, when we conider only active neutrino masses, as in Fig.~\ref{fig:aaf_DM_DH_rd_panels}, the total neutrino mass is varied as $\Sigma m_\nu$ between $0$ and $10$~eV. For the cases considering sterile neutrinos, we fix the total active neutrino mass to 0.058 eV (modeled as a single mass state), and vary $m^{\textrm{eff}}_{\textrm{s}}$ and $\beta$ so that the extra contribution to the effective number of relativistic species, $ N_{\rm eff}$, is between $0$ and $100$, and the sterile neutrino mass $m_s$ from $0.0001$ to $3$~eV.

\section{Results}
\label{sec:results}

Let us first consider our datasets without SH0ES' measurement of $H_0$. In this case, we can use Bayesian methods to determine the value of $\Neff$, with P18, CMB Lensing, DESI2 \& SNe PP datasets, assuming any extra relativistic energy density is massless,
\begin{equation}
\Neff = 3.10 \pm 0.16\, .
\end{equation}
Including SH0ES' measurement of $H_0$, with P18, CMB Lensing, DESI2 \& SNe PP datasets
\begin{equation}
\Neff = 3.43 \pm 0.13 \, ,
\end{equation}
which gives the best fit local expansion rate of this model to be $H_0 = 70.64\,\mathrm{km}\,\mathrm{s}^{-1}\,\mathrm{Mpc}^{-1}$. This reduces the tension between CMB-inferred $H_0$ and SH0ES's measurement from $5\sigma$ in $\Lambda$CDM to $2.4\sigma$ in a sterile neutrino cosmology, given SH0ES' value $H_0 = 73.04 \pm 1.04\, \mathrm{km}\,\mathrm{s}^{-1}\,\mathrm{Mpc}^{-1}$. Although not fully resolved, the Hubble tension is significantly reduced with larger $\Neff$ \cite{Escudero:2020ped,Allali:2024aiv}.
For the case of $\Neff$ being harbored in a massive sterile neutrino, we adopt $m_s = 0.1\,\mathrm{eV}$ as a representative mass, and we find 
\begin{equation}
 \Neff    = 3.50 \pm 0.13\, ,\label{eq:neffhigh}
\end{equation}
where all of these $\Neff$ values were determined by Bayesian methods. We tested the result in Eq~\eqref{eq:neffhigh} with a frequentist profile likelihood, and found consistent results. 

To explore the impact of fixing $N_{\rm eff}$ to a higher value motivated by $H_0$ data, we perform a profile likelihood analysis on $m_s$, adopting the massive sterile neutrino inferred best-fit value of $N_{\rm eff} = 3.50$. The sterile neutrino mass represents a parameter subject to a strict physical boundary, $m_s \geq 0$. In this case, the standard asymptotic assumptions underlying Wilks’ theorem are not valid, since the maximum-likelihood estimator and the likelihood ratio test statistic are no longer parabolically related when the unconstrained best fit lies in the unphysical region. Consequently, the naive use of $\Delta\chi^2$ contours can lead to under- or over-coverage near the boundary \cite{Herold:2024enb}.

\begin{figure}[t]
    \centering
    \includegraphics[width=\columnwidth]{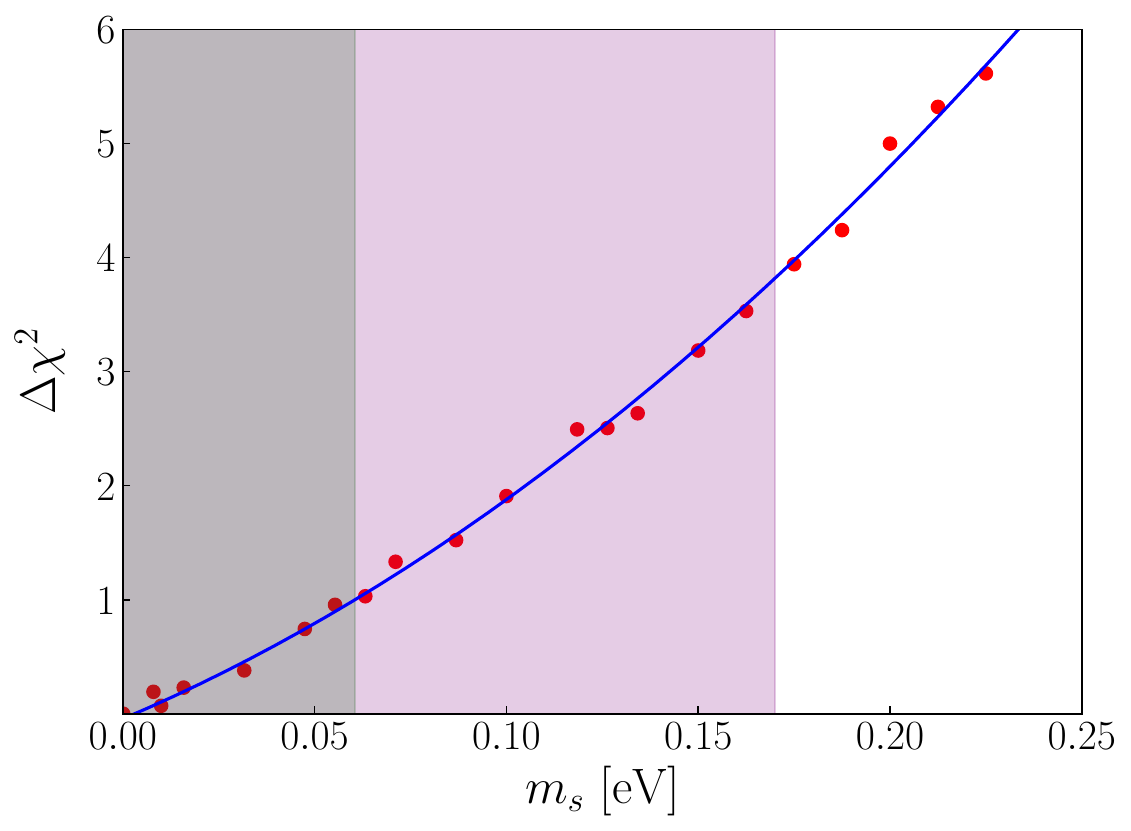}
    \caption{We show the profile likelihood for the physical sterile neutrino mass $m_s$, with the effective number of neutrino species fixed to $N_{\rm eff} = 3.50$. The red points show the computed $\Delta \chi^2$ values at discrete mass points used in the fit, while the blue curve represents a parabolic interpolation through these points. Shaded bands indicate the 68\% (black) and 95\% (purple) confidence intervals, respectively.}
    \label{fig:proflike_ms}
\end{figure}

To address the physical boundary, we follow the Feldman--Cousins prescription~\cite{Feldman:1997qc}, performing the profile likelihood scan in $m_s$ by maximizing over cosmological and nuisance parameters at each fixed $m_s \geq 0$, as shown in Fig.~\ref{fig:proflike_ms}. The resulting $\Delta\chi^2(m_s)$ profile was then fit with a quadratic function restricted to the physical branch. The 68\% and 95\% CL intervals were then derived from the intersection of the observed profile with the modified critical values. For more details on this method, see Ref.~\cite{Herold:2024enb}. For our data, this construction yields an upper limit of 
\begin{equation}
m_s < 0.170 \,\mathrm{eV} \ (95\%\,\mathrm{CL)}\, .
\label{eq:mslimit}
\end{equation}
Since in our $\Lambda$CDM model the active neutrino mass is fixed to a single massive state with $m_\nu = 0.058 \,\mathrm{eV}$, the upper bound on the total neutrino mass, $\Sigma m_\nu$, including both active and sterile species, is relaxed by a factor of 4.3 compared to the standard constraint of $\Sigma m_\nu < 0.053\,\mathrm{eV}$ (95\% CL) obtained from a similar dataset that does not attribute the extra radiation energy density to a sterile neutrino \cite{DESI:2025ejh}. The relaxation arises from the physics discussed in \S\ref{sec:implicat}. For a fixed $\Neff$, we find that Bayesian methods find a consistent limit for $m_s$ as in Eq.~\eqref{eq:mslimit}.

To explore the preference of SH0ES $H_0$ for sterile neutrinos as a function of exact $m_s$, as compared to the baseline $\Lambda$CDM case with $N_{\rm eff}=3.044$, where $\Delta N_\mathrm{eff}=0$ and $m_\mathrm{s}$ undefined, as no additional sterile neutrino is present, we find 
\begin{align}
m_s &= 0.1\,\mathrm{eV}, \quad \Delta \chi^2 = -10.1 \nonumber \\ 
m_s &= 0.2\,\mathrm{eV}, \quad \Delta \chi^2 = -7.0\nonumber \\
m_s &= 0.3\,\mathrm{eV}, \quad  \Delta \chi^2 = -3.2\nonumber \\
m_s &= 1\,\mathrm{eV}, \phantom{.0} \quad \Delta \chi^2 = 34.6\, ,
\label{eq:mschisq}
\end{align}
where negative values of $\Delta\chi^2$ correspond to an improved fit relative to $\Lambda$CDM. Since both $N_{\rm eff}$ and $m_s$ are fixed in all minimizations, the compared models have the same number of free parameters. The lightest masses are the most preferred, with even the case of $m_s = 0.1$ eV favored at greater than $3\sigma$.

We also tested whether the inclusion of the full-shape galaxy power spectrum data, together with the BAO feature, affects the constraints on $m_s$. Using SDSSBAOFS in combination with our baseline dataset in a Bayesian analysis—excluding DESI2 to avoid overlap with SDSS and because the DESI full-shape data is not yet publicly available—we find that the constraints become less stringent. For $\Neff=3.5$, the limit is $m_\mathrm{s} < 0.238$ eV (95\% CL) with the P18+SDSSBAOFS+PP+$H_0$ dataset, weaker than from BAO alone.

\section{Discussion \& Conclusions}
\label{sec:discussion}

In this work, we show how BAO observables are less sensitive to $\Neff$ and $\Sigma m_\nu$ together, than when either alone is left free, due to inherent degeneracies in the geometric observables of BAO. Though the Hubble tension may be alleviated from increasing $\Neff$ alone \cite{Riess:2016jrr}, we find that evidence for extra relativistic energy density in the combination of CMB and local expansion rate measurements, relaxes the limit on the mass of a partially thermalized sterile neutrino, and is greater than a factor of 4 weaker than standard massive neutrino limits, Eq.~\eqref{eq:mslimit}. Moreover, cosmologies with partially thermalized sterile neutrinos ($\Neff = 3.5$), for $m_s \ll 1\,\mathrm{eV}$, are preferred at $3.2\sigma$ [Eq.~\eqref{eq:mschisq}] over $\Lambda$CDM. Our updated analysis here, with DESI2, reinforces our earlier findings \cite{Escudero:2024uea} that sub-eV sterile neutrinos are preferred by the tension data. This reflects previous hints for extra $\Neff$ and sterile neutrinos, as discussed in the introduction. We have also found that Bayesian methods are problematic for mapping the parameter space for sterile neutrino mass when $\Neff$ is free, as the light sterile neutrino mass can increase to the warm dark matter (WDM) keV scale, as discussed in the Appendix. These prior volume effects may have impacts on WDM constraints that use $1/m_s$ mappings \cite{Irsic:2017ixq,Garzilli:2019qki,Villasenor:2022aiy}.

There is also an important connection to hints of light, eV-scale sterile neutrinos as responsible for short-baseline neutrino oscillation experiment results from LSND and MiniBOONE \cite{LSND:2001aii,MiniBooNE:2010idf}, anomalies observed in Gallium experiments SAGE \cite{Abdurashitov:2005tb}, GALLEX \cite{Kaether:2010ag}, and BEST \cite{Barinov:2021asz,Barinov:2022wfh}, as well as $\nu_\mu$ disappearance in ICECUBE \cite{IceCubeCollaboration:2024nle}. These experiments may suggest the presence of sterile neutrinos with mass-squared differences with the active neutrinos of order 1~eV$^2$.  Importantly, recent results from MicroBooNE show no evidence for an anomalous excess of electronlike events and exclude an electronlike interpretation of the MiniBooNE low-energy excess at $>99\%$ C.L.~\cite{MicroBooNECollaboration:2024cpi}. In order to accommodate constraints from other short-baseline experiments, models with more than one sterile neutrino have been considered, with the extra neutrino mass eigenstates having even larger mass differences than 1 eV$^2$ \cite{Conrad:2013mka}, and the effects of statistical methods have been investigated in detail regarding combined experiments \cite{Villarreal:2025mhv,Villarreal:2025ged}.

Our results disfavor light sterile neutrinos above the $\sim$0.3 eV scale. Therefore, there is tension with the mass scales preferred by short-baseline oscillation experiments. However, it is well known that cosmological neutrino mass constraints are alleviated in cosmologies with relaxed assumptions on dark energy, curvature, and other often-fixed cosmological parameters (e.g., see Refs.~\cite{Brinckmann:2020bcn,Elbers:2024sha}). Relaxation of models of reionization can also alleviate neutrino mass constraints \cite{Jhaveri:2025neg}. The subsequent reanalysis of Planck primary CMB and lensing data by Ref.~\cite{Tristram:2023haj} finds slightly weaker bounds on massive neutrinos \cite{Escudero:2024uea}, which could also further accommodate massive sterile neutrinos. Exploring multiple extra-parameter models, dynamical dark energy, and more complex models of reionization are beyond the scope of our present work, but are of interest if the Hubble tension and evidence for short-baseline oscillations persist. Full-shape measurements of cosmological matter clustering may enhance constraints on the presence of massive neutrinos, but our analysis finds they are not yet significantly constraining given the available full-shape dataset from SDSS DR12, as discussed above. 

Importantly, we find that a sterile neutrino cosmology should have a contribution to the relativistic energy density at below that of a fully-thermalized sterile neutrino ($\Delta \Neff = 1$). Such a scenario is inconsistent with the mixing angles required for the short-baseline results, since they would thermalize the sterile neutrino \cite{Langacker:1989sv}. However, partial thermalization can arise from models that suppress thermalization, including low-reheating temperature universes and models with lepton asymmetries \cite{Abazajian:2002bj,Jacques:2013xr,Hasegawa:2020ctq}. These cosmologies can also accommodate constraints from BBN. Interestingly, recent CMB results from ACT DR6 find $\Neff = 2.86 \pm 0.13$, which increases the tension between CMB datasets and $\Neff$ solutions to the Hubble tension \cite{ACT:2025tim}. 

The $H_0$ tension has motivated renewed attention to the role of neutrinos as messengers of possible new physics, especially as current generation surveys such as DESI and Euclid will deliver increasingly precise BAO and LSS measurements across a broad redshift range with a wide variety of statistics \cite{DESI:2024mwx,Euclid:2024yrr,Escudero:2022rbq}. Future CMB surveys will also have increased sensitivity to $\Neff$ \cite{CMB-S4:2016ple,Green:2019glg,SimonsObservatory:2025wwn}.
We find that cosmological data can be consistent with, and may even prefer, short-baseline-oscillation motivated light sterile neutrinos, though with mass scales that tend to be lighter than that preferred by the oscillation experiments. As further oscillation and cosmological data is available, we will uncover whether the elusive neutrino sector is responsible for both oscillation and cosmological tensions.

\begin{acknowledgments}
HGE and KNA would like to thank Laura Herold for her detailed comments on the manuscript. We thank useful discussions with  Julien Froustey,  George Fuller, Kevin Huffenberger, Ryan Keeley, Jay Krishnan, Antony Lewis, Bill Louis, Levon Pogosian, Georg Raffelt, Michael Ryan, and Cannon Vogel. KNA would like to thank the organizers of the January 2025 Neutrinos in Physics and Astrophysics Workshop: in Celebration of the Careers of George Fuller and Baha Balantekin, where many of these discussions took place. KNA is partially supported by the U.S. National Science Foundation (NSF) Theoretical Physics Program Grant No.\ PHY-2210283. 
\end{acknowledgments}

\appendix*

\section{Prior Dependence and Volume Effects}
\label{priordep}

\begin{figure}[t]
    \centering
    \includegraphics[width=\columnwidth]{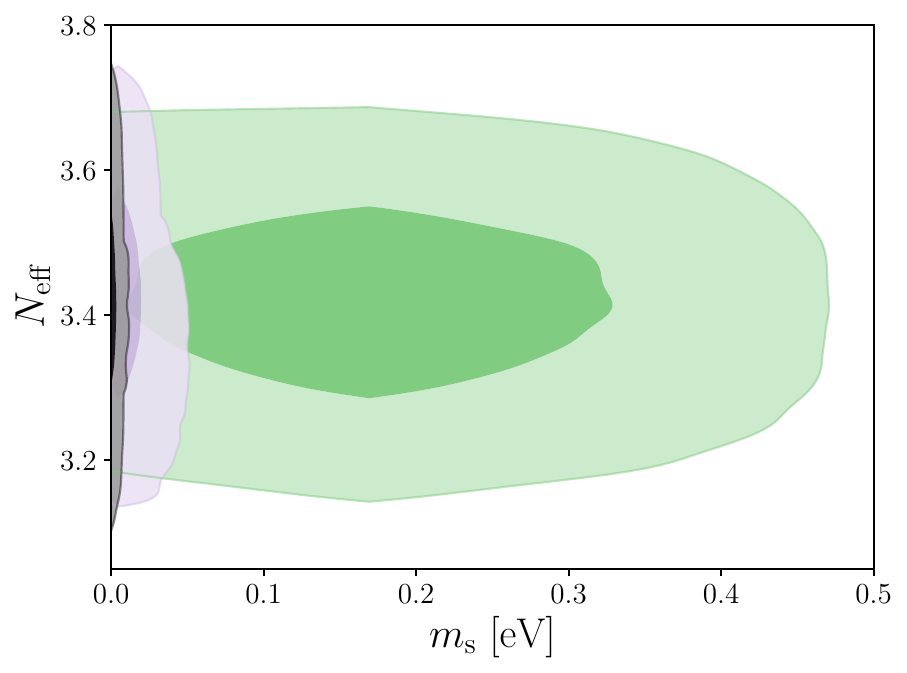}
    \caption{
    Two-dimensional posterior distributions are shown for the physical sterile neutrino mass, $m_{\rm s}$ and the effective number of neutrino species, $N_{\rm eff}$, under three different prior choices using MCMC Bayesian methods and our baseline datasets. The black contours correspond to a prior set as $1/\sqrt{m_s}$, the light purple contours to a prior as $1/m_s$, and the pale green contours, which give the most relaxed bounds, correspond to a prior of $1/m_s^2$. This figure illustrates the impact of prior volume effects in Bayesian analyses of this cosmological model.}
    \label{fig:prior_effects_mnu_neff}
\end{figure}

Bayesian inference is sensitive not only to the likelihood function but also to the choice of priors. When placing constraints on the sterile neutrino mass $m_s$, it is therefore essential to carefully consider how priors are defined, particularly under non-linear reparametrizations of the mass. 
A notable bias arises when a uniform prior is imposed not on $m_s$ itself, but on transformed variables such as $1/m_s$. While such choices may be motivated by wanting to map an infinite parameter space where $m_s$ can become a heavy dark matter particle, the inverted parameter does not correspond to a uniform weighting in $m_s$ space. Instead, it redistributes the prior volume in a non-trivial way and can significantly bias the resulting posterior constraints.

One can test the volume dependence by using inverted weighted scalings, $1/m_s$, $1/m_s^2$, or $1/\sqrt{m_s}$, where we use the standard posterior distribution estimation package \texttt{getdist} \cite{Lewis:2019xzd}. For example, imposing a flat prior on $1/m_s$ induces an effective prior on $m_s$ proportional to $1/m_s^2$, thereby strongly favoring small values of the mass. Likewise, a flat prior on $1/\sqrt{m_s}$ produces an induced prior scaling as $1/m_s^{3/2}$, and a flat prior on $1/m_s^2$ leads to an even steeper weighting. In all of these cases, the prior volume is disproportionately concentrated at low $m_s$, independent of the likelihood, artificially driving the posterior distribution toward vanishing masses. This behavior is problematic when interpreting constraints on sterile neutrino masses, as shown in Fig.~\ref{fig:prior_effects_mnu_neff}. For example, we find that when using a flat prior on $1/\sqrt{m_s}$ in DESI2 analyses, the posterior is biased toward lower masses, even in regions where the likelihood is relatively flat.  Our inferred upper limits on the sterile neutrino mass vary widely with the choice of prior. For example, a prior uniform in $1/\sqrt{m_s}$ gives the tightest limit, $1/m_s$ yields a bound roughly five times larger, and $1/m_s^2$ gives the loosest constraint, over twenty times the tightest. This illustrates that the derived limits on $m_s$ can be driven primarily by the prior rather than the data itself.

In general, when you place a flat prior on a transformed parameter, you are not placing a flat prior on $m_s$ itself. Instead, you are reweighting the space of $m_s$ values, giving disproportionate volume to certain ranges—typically the low-mass region. This violates the principle of non-informative priors unless the transformation is physically motivated and explicitly accounted for in interpretation. This issue may be worth considering when assessing cosmological limits on WDM particle masses or light sterile neutrinos, where the physical quantity of interest is $m_s$ itself \cite{Irsic:2017ixq,Garzilli:2019qki,Villasenor:2022aiy}. 

Using transformed priors in such contexts can severely misrepresent the viable parameter space and overstate the level of constraint. Flat priors on $m_s$ avoid such distortions when the mass is the physical quantity of interest. This approach ensures that each interval in $m_s$ contributes equally to the prior volume and that the resulting posterior more faithfully reflects the data's constraining power, which can also be achieved by frequentist methods.

\bibliography{references}

\end{document}